\documentclass[useAMS,usedcolumn]{mn2e}
\pdfoutput=1 
\usepackage{rotating}
\usepackage{lscape}
\usepackage{latexsym}
\usepackage{graphicx}
\usepackage{sidecap}
\usepackage{acronym}
\usepackage{subfigure}
\usepackage{textcomp}
\usepackage{bbding}
\usepackage{epsf}
\usepackage{amsmath,amssymb}
\usepackage{float}
\usepackage{multirow}
\usepackage{flafter}
\usepackage{placeins}
\usepackage{afterpage}
\usepackage{preview}
\usepackage{here}
\usepackage{color}

\newcommand{\OIII}{[O\,{\sc iii}]}
\newcommand{\OII}{[O\,{\sc ii}]}

\def\p0{\phantom{0}}

\def\cm3{cm$^{-3}$}

\def\12{$^{12}$CO}
\def\13co{$^{13}$CO}

\def\la{{\hbox{$\lesssim$}}}

\def\arcdeg{\hbox{$^\circ$} }
\def\arcmin{\hbox{$^\prime$} }
\def\arcsec{\hbox{$^{\prime\prime}$} }
\def\ergcms{erg cm$^{-2}$ s$^{-1}$}

\title[A bipolar planetary in open cluster ESO\,96-SC04]{PHR1315--6555: a bipolar  planetary nebula in the compact Hyades-age open cluster ESO\,96-SC04}

\author[Quentin A. Parker et al.]
{Quentin A. Parker$^{1,2}$\thanks{E-mail: qap@ics.mq.edu.au}, David J. Frew$^{1}$,  Brent Miszalski$^{3,1,5}$, Anna V. Kovacevic$^{1}$, \newauthor Peter M.  Frinchaboy$^{4}$, 
Paul D. Dobbie$^{2}$, and Joachim K\"oppen$^{5,6,7}$\\ 
  $^1$ Department of Physics \& Astronomy, Macquarie University, North Ryde, NSW 2109, Australia \\
  $^2$ Australian Astronomical Observatory, PO Box 296, Epping, NSW 1710, Australia\\
  $^3$ Centre for Astrophysics Research, STRI, University of Hertfordshire, Hatfield, AL10 9AB, UK\\
  $^4$ Department of Physics \& Astronomy, Texas Christian University,  2800 South University Dr, Fort Worth, TX, 76129, USA\\
$^5$ Observatoire de Strasbourg, 11 Rue de l'Universit\'e, F--67000
  Strasbourg, France\\
  $^6$ International Space University, Parc d'Innovation, 1 Rue
  Jean-Dominique Cassini, F--67400 Illkirch--Graffenstaden, France\\
  $^7$ Institut f\"ur Theoretische Physik und Astrophysik, Universit\"at
  Kiel, D--24098 Kiel, Germany  }

\date{Accepted 
      Received 
      in original form }

\begin{document}

\maketitle 
\begin{abstract} 
We present a detailed study of a bipolar,  possible Type~I  planetary nebula (PN), PHR1315-6555 (PN G305.3-03.1), that was discovered as part of the Macquarie/AAO/Strasbourg H$\alpha$ planetary nebula project (MASH) and that we considered at the time was an excellent candidate for membership of the distant, compact, intermediate-age open cluster, ESO\,96-SC04.  The strong evidence for this association is presented here making this the only known example of a PN physically associated with a Galactic open cluster. Cluster membership is extremely important as  it allows for very precise estimates of the fundamental properties of the PN as the cluster is at a known distance. 

The PN was discovered by one of us (QAP) during systematic MASH searches for new Galactic PNe of the AAO/UKST H$\alpha$ survey  and had been missed in earlier broadband surveys, including specific CCD studies of the host cluster.  We present original discovery images and CTIO 4m MOSAIC-II camera follow-up narrow-band images that  reveal its bipolar morphology. We also present: (i) low-resolution optical spectra that spectroscopically confirm the PN; (ii) accurate radial velocities of PN and cluster stars from high resolution spectroscopy which show they are consistent; and (iii) a reliable, independent distance estimate to the PN  using a robust  PN distance indicator which agrees with the published cluster distance to within the errors. We also provide preliminary estimates of basic PN properties  and abundance estimates from deeper spectra that show it to be of possible Type~I chemistry. This is also consistent its estimated turn-off mass. Taken together these findings present a powerful case for clear physical association between the PN and host cluster. Results for this  association will be of considerable  interest to specialists across differing astrophysical disciplines, including PNe, white dwarfs, and open clusters. 
\end{abstract}

\begin{keywords}
Stars: Planetary Nebulae; PN G305.3-03.1; Open clusters, White Dwarfs
\end{keywords}

\section{Introduction}
An association between a planetary nebula (PN) and an open star cluster is a very valuable astrophysical tool.  This is because the accurate cluster distance, determined from a colour-magnitude diagram (CMD),  constrains the physical parameters of the PN and central star (CSPN) to exceptional precision.  The age and  mass of the progenitor star can be tightly constrained from theoretical cluster isochrones, while CSPN photometry allows a precise determination of its intrinsic luminosity and mass. The progenitor star mass, which can be related to the chemistry of the resulting PN (from spectroscopy), provides a rare additional datum for the fundamental white dwarf (WD) initial-to-final mass relation  (IFMR) currently best determined from cluster white dwarfs (e.g. Williams et al. 2004; Ferrario et al. 2005;  Kalirai et al. 2008; Dobbie et al. 2009) which intimately links WD properties to their main-sequence progenitors.  A robust IFMR is a key component of using WD luminosity functions to constrain the age of the Galactic disk (using the field WD population) and open clusters (using the cluster population) and is also key to  mapping the build up of carbon and nitrogen in galaxies.  Note that measuring precise stellar masses from PN is also possible, if somewhat controversial -- see the discussion in Gesicki \& Zijlstra (2007).

Unfortunately, the number of PNe that are genuine members of Galactic star clusters  of all kinds is extremely small (there are four currently known in Galactic globular clusters; in M15, M22, Pal 6 and NGC 6441; Jacoby et al. 1997).  Our discovery (Parker et al. 2006) of a faint bipolar PN (PHR1315-6555) within 23~arcseconds of the projected centre of the distant,  compact, intermediate-age Galactic open cluster ESO 96-SC04, is therefore of great interest.\footnote{ At least two PN candidates have also been found in globular clusters in external galaxies, e.g. Larsen (2008), for a peculiar PN candidate in a globular cluster in the Fornax dwarf spheroidal galaxy and Minniti \& Rejkuba (2002) for a PN candidate in globular cluster G169 in Centaurus A (NGC5128).}

Many other PN-cluster candidates  (e.g. Bonatto, Bica \& Santos 2008) have been shown to be either likely or definite line-of-sight superpositions. For example Majaess, Turner \& Lane (2007) showed that 50 per cent of known possible PN-cluster associations were spatial coincidences, while 
Frew (2008)  showed that  a case against association can be made for nearly all previous implied PN-cluster pairings. Indeed, one of  the better candidates for a PN-cluster association, that  between the PN NGC~2438 and the open cluster M~46  (e.g. Pauls \& Kohoutek 1996;  Majaess, Turner \& Lane 2007;  Bonatto, Bica \& Santos 2008) has recently been eliminated via a detailed radial velocity study of  the PN and large numbers of cluster stars (Kiss et al. 2008). 

In this paper we present preliminary discovery and detailed subsequent follow-up data on our remarkable find and its host cluster. These are used to build a compelling and robust case for the veracity of the PN-cluster association. 

\section{Discovery and spectroscopic confirmation of PN PHR1315-6555}
The discovery of a PN  in the open cluster ESO\,96-SC04  was made via images from the AAO/UKST 
H$\alpha$ survey (Parker et al. 2005). This survey  has produced a  map of ionised gaseous emission with arcsecond resolution and 2-5~Rayleigh sensitivity over 4000 square degrees of the southern Galactic plane. 
Both the original 3-hour H$\alpha$ survey film exposures and matching short 15-min R-band 'SR' film exposures (which reach almost identical  depth for continuum sources) have been scanned with the SuperCOSMOS measuring machine (e.g. Hambly et al. 2001) at the Royal Observatory 
Edinburgh as part of the SuperCOSMOS H$\alpha$ Survey (SHS) and are available on-line.\footnote{The 0.67~arcsecond ($10\mu$m)  pixel 
H$\alpha$ and matching SR data can be obtained here: http://www-wfau.roe.ac.uk/sss/halpha/} The object was found in H$\alpha$ 
survey field HA~137 by one of us (QAP) during searches for new Galactic PNe as part of the MASH PN project  (Parker et al. 2006, Miszalski et al. 2008a). MASH includes data for over 1200 spectroscopically confirmed new PNe, nearly doubling the Galactic PNe population contained in the Acker (1992, 1996) catalogues. The PN has the MASH designation PHR1315-6555 (and IAU designation PN G305.3-03.1).

Despite the cluster being imaged  in $B$ and $V$ on the ESO NTT 3.5-m telescope by Carraro et al. (1995) the PN was not noted in these data
due to its low-surface brightness and small apparent size ($\sim$20~arcseconds). We have now also taken deep CTIO 4-m MOSAIC-II camera  images of the cluster which confirm its compact nature with an effective cluster diameter of $\sim$80~arcseconds. These consisted of narrow-band [O~III] and H$\alpha$+[N~II]  on and off-band frames  that  also clearly reveal the bipolar PN  morphology (see Fig.~\ref{PNimages}) and show the PN has an overall size of  
$18\times14$~arcseconds.  The CTIO observations were made on 11$^{\rm th}$ June 2008 during a 
separate mission to obtain [O~III] imaging for a large sample of Galactic Bulge PNe for accurate flux determinations (Kovacevic et al. 2010). 
Unfortunately we have been unable to unequivocally identify the CSPN in our current data due to its estimated very faint magnitude (see Sec.6).

A preliminary, low resolution confirmatory spectrum was obtained for PHR1315-6555 at the 1.9-m telescope of the
South African Astronomical Observatory (SAAO) in May 2001 shortly after its initial discovery. The slit was 
$\sim$2.4~arcseconds  wide and  was fixed east-west.  A 300 lines/mm low dispersion grating was used giving a spectral coverage of 
3300--7300\AA. Spectrophotometric  standard stars from Stone \& Baldwin (1983) were observed  to 
flux calibrate the spectrum which were reduced and analysed using standard IRAF routines. The presence of  a strong He~II 4686\AA~ emission line
indicated a high excitation nebula. 
Medium resolution SAAO spectra were subsequently obtained in  February 2004 during general MASH  observing runs in an attempt to obtain an accurate radial velocity for this PN.  A 1200 line/mm grating was used blazed for the red spectral range.  
Unfortunately, the modest resolution of these spectra were still not sufficient to provide the very accurate determination 
of the PN radial velocity necessary to tie in precisely to the open cluster velocity.  However,  these red data do separate 
the [N~II] and H$\alpha$  lines more clearly which confirm a high [N~II]/H$\alpha$ ratio of  $\sim$2.0. Such high values 
are typical of many  Type~I bipolar PN (Peimbert \& Torres-Peimbert 1983; Perinotto \& Corradi 1998; Frew \& Parker 2010). Deep, high S/N, medium resolution spectra were  then taken with the ANU 2.3m double beam spectrograph  (DBS; Rodgers et al. 1988) in May 2006 to try to measure the abundances of the PN. The spliced  high S/N blue and red spectra from this observation are shown in Fig.~\ref{PN-spec} which demonstrates  typical PN emission lines and ratios and is sufficiently deep to detect key diagnostic nebular lines for nebula electron density and temperature determinations (see Sec.5.). Note great care is taken in splicing the DBS flux calibrated red and blue arms to that reliable Balmer decrement values can be obtained. Kovacevic et al. (2010) compared the spliced DBS $H\alpha/H\beta$ ratios to those collated from the literature, where they exist. The distribution of the comparisons, of which there are 14, has a standard deviation of 0.04dex, or 10 per cent, giving us confidence in our splicing process for these spectra.

\section{Discovery of the host cluster ESO\,96-SC04}
ESO\,96-SC04 (AL\,1) is an interesting, compact open cluster first recorded by Andrews \& Lindsay (1967) on ADH Schmidt plates, with a reported angular diameter of 75~arcseconds. It was independently discovered by van den Bergh \& Hagen (1975) as BH\,144 before being noted on ESO Schmidt plates and designated ESO\,96-SC04 (Lauberts 1982). 

There has been much confusion in the literature regarding the nomenclature of this cluster, a consequence of a number of published positions being in error. The current SIMBAD values are, however, accurate.
We  also carefully measured the cluster's position from the SHS short-red pixel data (Parker et al. 2005) which have a  world coordinate system (WCS) accurate to $\sim$0.3~arcseconds in-built. This new determination is $13^{\rm h}15^{\rm m}16^{\rm s}$ $-65^{\arcdeg} 55^{\arcmin} 16^{\arcsec}$ (J2000)
which is in excellent agreement with that  determined by Carraro et al. (2005) and with the position previously published by Lauberts (1982).  Note the positions given by Tadross (2001) and Tadross et al. (2002) are in error.

\begin{table}
\begin{center}
\caption{Fundamental literature parameters for ESO\,96-SC04}
\medskip
\label{summary}
\begin{tabular}{llccc}
\hline
Reference      	& Telescope    	&   	D   		& 	$E(B-V)$ 		& Age \\
     			&    			 &   (kpc)   		& 				& (Myr)  \\ \hline                                                            
PJM94, JP94    &  0.9m CTIO 	&	7.57  	&     0.72           		&     ...           \\
CVO95            	&  3.5m NTT 	&	11.8   	&     0.75    		&   700         \\
CM04          	&  1.0m SAAO  &12\,$\pm$\,1   	&     0.7\,$\pm$\,0.2   &   800         \\
CJE05            	&  1.0m CTIO 	&10.1*   		&     0.7*     		&   800         \\
\hline
Adopted values & & 10.4$\pm$1.8 & 0.72 & 800\\	                                                                   
\hline
\end{tabular}
\end{center}
References:  CVO95, Carraro et al. (1995); PJM94, Phelps et al. (1994); JP94, Janes \& Phelps (1994); CM04, Carraro \& Munari (2004); CJE05, Carraro et al. (2005). * Corrected  values -- see text for details.
\end{table}

\section{PN membership of open cluster ESO\,96-SC04}

Our strong assertion of PN cluster membership  is based on three key  arguments and several other contributing pieces of corroborating evidence. 
First is the very close (23~arcsecond) angular proximity
of the PN to our newly determined central position for the cluster ESO\,96-SC04. No other purported association has such a close angular proximity  to the host cluster core. Secondly, we have excellent radial velocity agreement  of the PN and cluster stars to within 1~km~s$^{-1}$. This is important as open cluster velocity dispersions are typically only $\sim$1~kms$^{-1}$; e.g. see the extensive series of WIYN open cluster radial velocity studies such as Mathieu (2000) and Hole et al. (2009). Thirdly, there is very good agreement between our independently estimated PN distance (from our new surface brightness radius relation- see  sec.4.3.2 below)  and that of the host cluster to within the errors. Reddening plays an important role in these distance estimates and again independently determined cluster and PN reddening estimates are in good agreement. 
Together these  main points provide the strongest evidence for association. There are other, more circumstantial and/or less well established pieces of evidence that also support association  such as  Galactic scale-height arguments and likely progenitor mass and these are briefly addressed later.

\subsection{Close angular proximity of PN and cluster}
\begin{figure*}
\includegraphics[width=16.5cm,height=5.5cm]{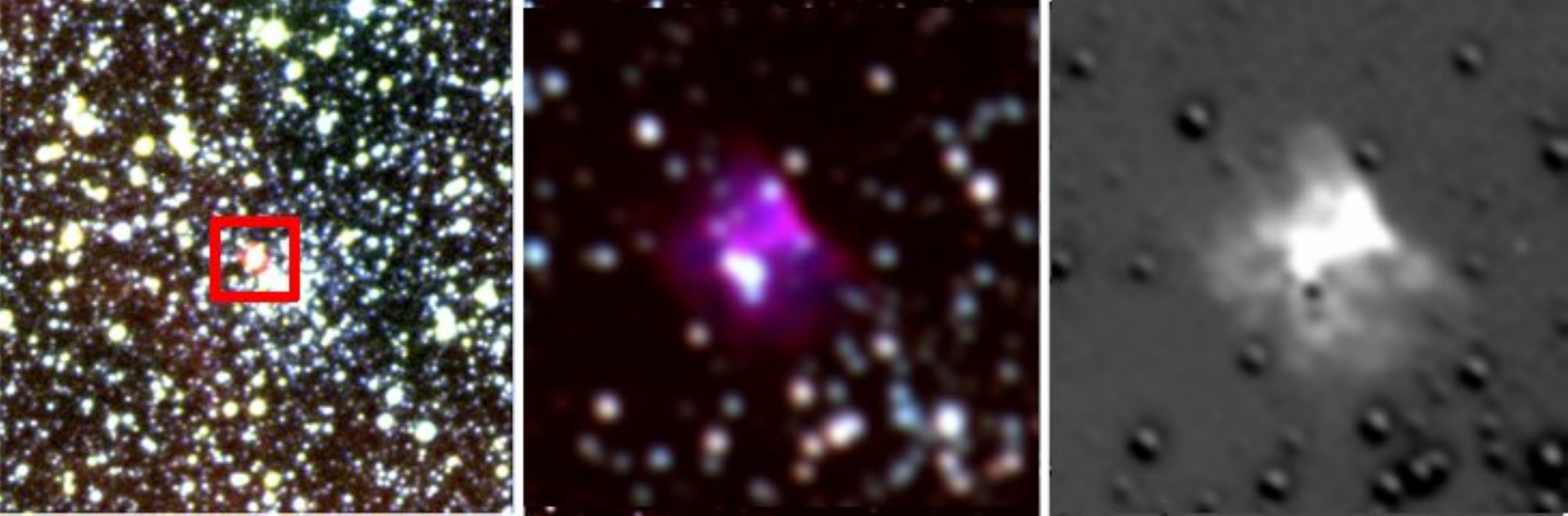}
\caption{Left panel is a $5\times5$~arcminute colour composite SuperCOSMOS image comprising the SuperCOSMOS H$\alpha$ 
image (red), matching broad-band SR image (green) and B${_J}$-band image (blue). The image is centred on the PN which falls well 
within the half-light radius of the cluster. The middle panel shows a $1\times1$~arcmin CTIO 4-m MOSAIC-II camera colour composite centred on the PN  with its  bipolar morphology evident. The montage was created from the H$\alpha+$[N~II] (red), [O~III]-off (green) and [O~III] (blue) narrow-band filters. The right panel of the same area is a quotient image obtained by dividing the H$\alpha+$[N~II] image by the H$\alpha$ off-band
image which is effective in showing the fainter outer regions of this unique bipolar PN. North-east is to the top left in all panels.}
\label{PNimages}
\end{figure*}

In Fig.~\ref{PNimages} left panel we show the PN discovery images combined as a colour composite of the SuperCOSMOS H$\alpha$ image 
(red) and its matching broad-band SR image (green) --- Parker et al. (2005), and standard SuperCOSMOS Sky Survey  B${_j}$-band image (blue) 
--- Hambly et al. (2001). The image is $5\times5$ arcminutes in size with north-east to the top left. The accurate position of the PN determined from the SHS image data is $13^{\rm h}15^{\rm m}18.9^{\rm s}$ $-65^{\arcdeg}55^{\arcmin}01^{\arcsec}$.
Note  the compactness of this cluster which has an apparent diameter of only $\sim$80~arcseconds.  Even in this lower resolution data the PN appears to be a  bipolar object though the precise morphology is difficult to ascertain. The middle  panel shows a $1\times1$~arcmin higher angular resolution (0.5~arcseconds/pixel) CTIO 4-m MOSAIC-II camera colour composite image of the open cluster with the bipolar  morphology now  clearly evident.  This image was created from  H$\alpha+$[N~II] (red), [O~III]-off (green) and [O~III] (blue) narrow-band filters\footnote{http://www.noao.edu/kpno/mosaic/filters/}. The right panel of the same area is a quotient image obtained by dividing the H$\alpha+$[N~II] image by the H$\alpha$ off-band image which is effective in showing the fainter outer regions of this  bipolar PN. The PN lies 23~arcseconds from the cluster centre as 
measured from the CTIO images, well within the estimated cluster half-light radius of  $\sim$32~arcseconds. 

\begin{figure*}
\includegraphics[width=18cm,height=12cm]{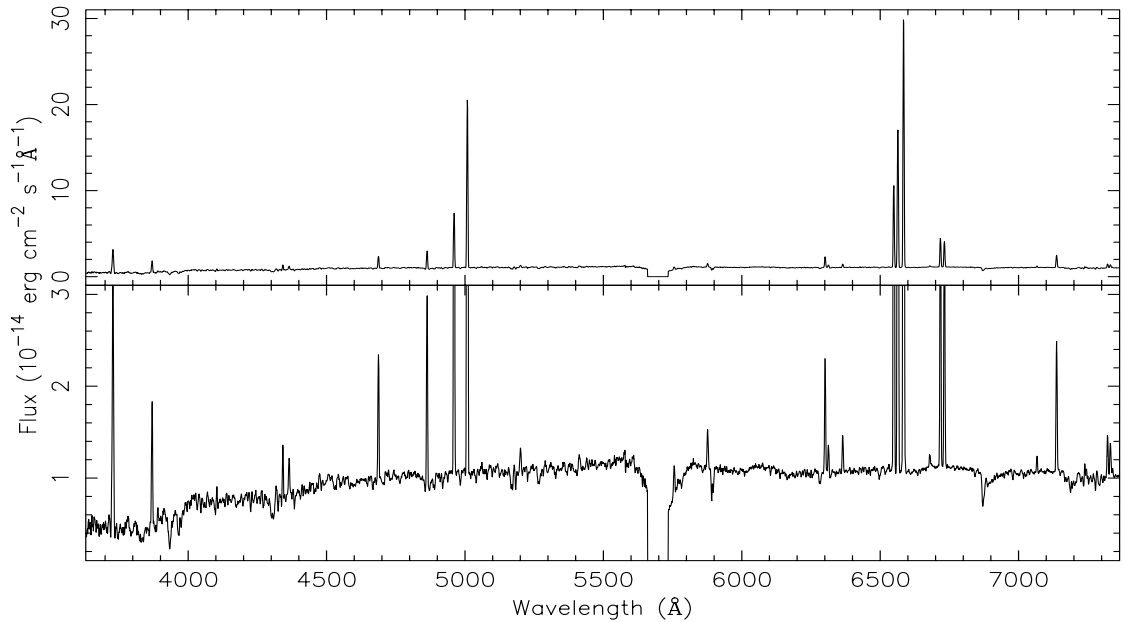}
\caption {Flux calibrated, medium resolution  2.3m DBS spectrum of PHR1315-6555 (PN G305.3-03.1) taken in May 2006. The
decent blue S/N means it was possible to establish the true strength of the  \OII\ 3727\AA\ doublet. The red and blue arm data have been independently flux calibrated (note gap at 5800\AA). The upper panel shows the spectrum with the full range of line intensities plotted while the lower panel is re-scaled to reveal the important, fainter diagnostic lines more clearly.
The strong He~II  and \OIII\ lines relative to H$\beta$ and the strong [N~II] relative to H$\alpha$ 
are clear signatures of a PN.  Also note that the high H$\gamma$ to [O~III] 4363\AA~ratio appears to resemble that seen in some symbiotic stars but that is clearly not the case here.} \label{PN-spec}
\end{figure*}

\subsection{Agreement of PN and cluster radial velocities}

Subsequent to the initial PN confirmatory SAAO spectrum in 2001 referred to earlier,
an 1800~second high-dispersion exposure  was obtained with the 2dF AAOmega multi-object spectrograph (e.g. Lewis et al. 2002; Saunders et al. 2004; Sharp et al. 2006) on the 3.9~m AAT on 29 May 2008 with VPH grating 1700D to  tie down the radial velocity of the PN to the required accuracy (note that due to the dual beam nature of this instrument both blue and red arm spectra are obtained). The far-red high dispersion spectrum of the PN was centred at 8600\AA~  and  provides a  quality radial velocity for the PN from the clearly  exhibited  Paschen series emission lines of hydrogen. 
The data were reduced using the AAO's 2dFDR software package to obtain accurate, 
corrected PN heliocentric velocity of 58~kms$^{-1}$ and a dispersion of only 2.5~kms$^{-1}$ from six Paschen lines (from P12-P19). 
This is in excellent agreement with the observations of three cluster stars in ESO 96-SC04 observed using the CTIO 4-m + Ritchey-Cretien spectrograph  in August 2004 covering the region 6400--8900\AA~ by one of us (PF).   These spectra were reduced using standard IRAF techniques and radial velocities determined using {\tt fxcor}, are listed in Table~\ref{starvels}. These yielded a cluster heliocentric velocity of 57 $\pm$ 5 km s$^{-1}$. Note the individual star identifications are from Janes \& Phelps (1994). We also provide photometry for these stars from Carraro et al. (2005) in Table~\ref{starvels}. 

\begin{table*}
\begin{center}
\caption{Summary of radial velocity measures for cluster stars from CTIO 4~m telescope spectroscopy together with broad band photometry from Cararro et al. (2005). Positions are in J2000 co-ordinates.}
\label{starvels}
\begin{tabular}{lcccccc}
\hline\hline
   ID & RA    & DEC   & $V_r$ (km s$^{-1}$) 		     &$B$	&	$V$	&	$I$ \\
\hline
   12718  &   13:15:16.01  &  -65:54:54.05 &   54 $\pm$ 7 &18.82	&	17.51	& 	15.48\\
   13152  &   13:15:19.31  &  -65:55:16.33 &   60 $\pm$ 7 &19.04	&	17.53	&	15.30\\
   13227  &   13:15:19.97  &  -65:54:33.89 &   57 $\pm$ 7 & 18.96	&	17.45	&	15.32\\
\hline
\end{tabular}
\end{center}
\end{table*}

The AAOmega blue arm data with the lower resolution 580V VPH grating 
gave a velocity of $59$~kms$^{-1}$ and $\sigma =2$~kms$^{-1}$, from the high S/N, strong [O~III] and H$\beta$ lines, which is still in excellent 
agreement with the higher resolution far-red spectrum. Open cluster velocity dispersions are typically of order $\sim$1~kms$^{-1}$  as already noted,  so this agreement is a very strong indicator that the PN is a cluster member, even though proper motion data are currently lacking.
A summary of our complete spectroscopic observations obtained for PHR1315-6555 is given in Table~\ref{specsum}.

While the excellent radial velocity agreement  between PN and cluster is a necessary condition for physical association  along this sight-line positive velocities of the order found are expected at these large distances beyond 9\,kpc so general radial velocity coincidence is not quite as strong a test as it might otherwise be.

\begin{table*}
\begin{center}
\caption{Summary of spectral observations  of  PN PHR1315-6555.}
\label{specsum}
\begin{tabular}{lllccl}
\hline\hline
Telescope  	&  Date observed    	& Dispersion 		& Wavelength coverage   				& Exposure   & Comment    \\
                            	&                  	       	& lines/mm     		&   (\AA)          				& (seconds)  &     \\ \hline
 1.9m SAAO      & 15/05/2001		& 300B      		&   3800--7800 				&  900		& First confirmatory spectrum    \\
 1.9m SAAO  	& 16/02/2004    	& 1200R     		& 6055--6815  				& 1200		& Radial velocity estimates \\
 2.3m MSSSO  &  21/05/2006             & 600B, 600R		& 3640--5650; 5730--7365	& 600, 3$\times$1200  &  Abundance estimates,  5" slit\\
 2.3m MSSSO   & 06-12 /05/2008  &  1200B, 1200R     	& 4100--5050; 6300--7250		& 300,  1200	& Wide 19" slit; PA=270/220$^\circ$ (300/1200s)\\
 3.9m AAT		& 29/05/2008   		& 580V, 1700D 	& 3740--5800, 8340--8800	& 1800 		& AAOmega for accurate radial velocity\\
\hline
\end{tabular}
\end{center}
\end{table*}

\subsection{Agreement of Cluster and PN  distances and reddening}
The third key element in our proof of PN cluster membership is the agreement between the independently determined cluster distance and the PN  
distance  established from our newly developed H$\alpha$ surface-brightness -- radius (SB-$r$) relation (Frew, Parker \& Russeil 2006; Frew \& Parker 2006) as discussed below. Accurate reddening estimates from both the cluster and PN are also required as these values fold into the distance calculations.

\subsubsection{Cluster reddening, distance and age estimates}
For the cluster a number of studies have been made over the last decade.  Janes \& Phelps (1994) estimated a preliminary 
distance of 7.57 kpc (see also Phelps et al. 1994 and Friel 1995), though Carraro et al. (1995) determined a larger distance of 11.8~kpc, and determined a reddening of $E(B-V)$ = 0.75.   Dutra \& Bica (2000) used DIRBE/IRAS 100$\mu$m dust emission to estimate a total line-of-sight reddening in the galactic disk of $E(B-V)$ = 0.94 in this direction.  For the cluster, they assumed a reddening of $E(B-V)$ = 0.72 for a distance of only  7.57~kpc, concluding that the difference of $E(B-V)$ = 0.22 is due to obscuration behind the cluster  so either the reddening determined from the CMD is underestimated, the total DIRBE reddening estimate is too high,  or there is significant dust at large distances from the galactic plane beyond the cluster.  The last alternative is the least likely.  Since the reddening of the PN estimated from our spectrophotometry is consistent with the various cluster CMD determinations (Janes \& Phelps 1994; Carraro et al. 1995; Carraro \& Munari 2004), the total DIRBE reddening estimate may be in error.

Note the formal Schlegel, Finkbeiner \& Davis (1998; SFD hereafter) asymptotic reddening along this sight line to the cluster is $E(B-V) = 0.93$ 
or c = 1.34. However, there is good evidence that the SFD values can be overestimated in higher extinction regions close to the plane by a factor of  $\sim$1.4, e.g. Yasuda et al. (2007).  Applying this correction leads to $E(B-V)$ = 0.66 which fits quite well with the published cluster reddening estimates above.  
Note Arce \& Goodman (1999) and Dutra et al. (2003, and references therein) also suggest the SFD extinction is overestimated at lower latitudes and at moderate/high values of reddening. All these results seem consistent with Yasuda et al. (2007) -- i.e. the SFD asymptotic reddening is 1.2 to 1.5 times too high in this region.

Carraro \& Munari (2004) derived an $E(B-V) = 0.7\pm0.2$ and a distance of 12.0~kpc to the cluster in very good agreement with their earlier estimates. Somewhat surprisingly Carraro, Janes \& Eastman (2005) estimated a distance of 16.9~kpc, which is considerably higher than previous determinations, due to their low adopted value for the reddening of $E(B-V)$ = 0.35.  We discount this later value which appears anomalous compared to all previous estimates and pertinently the reddening also derived for the PN. If a more likely value of $E(B-V)$=0.7 is adopted then cluster distance becomes 10.1~kpc, in much better agreement with their earlier determinations. 

Phelps et al. (1994) and Carraro et al. (2005) give an age of 800~Myr. The best fit solar abundance isochrone to the 
cluster CMD by Carraro et al. (2004) gives an age of $\sim700$~Myr or similar to the Hyades cluster, which translates to a progenitor 
mass of $\sim$2.5 M$_{\sun}$ (e.g. Girardi et al. 2000), which depends slightly on metallicity.    

The relevant cluster data are summarised later in Table~\ref{summary} and we adopt a cluster distance of 10.4~kpc averaged from the available literature data. The reader is also referred to the summaries of Dias et al. (2002) and Tadross et al. (2002). 

\subsubsection{ESO\,96-SC04 as seen in 2MASS}
In principle it is also possible to derive independent cluster reddenings from 2MASS JHK data of the cluster from late-K/M type members in a J-H versus H-K colour-colour plot due to the appearance of water vapour in the cooler atmospheres of late-K and M stars that produces a useful
inflexion in the stellar locus. Unfortunately, ESO 096-SC04 is too distant for accurate cluster photometry with 2MASS and the cluster is very hard to see in 2MASS data at all (see Parker et al. 2010). Only 36 2MASS stars are recorded across a 1~arcminute region centred on this rich compact cluster (recall half light radius is only 32~arcseconds) and only 20 have reliable photometry in all 3 2MASS bands.   Many of these are simply foreground stars. 
In fact, based on the JHK absolute magnitudes quoted by Kraus \& Hillenbrand (2007) IR photometry down to  K$\sim$20-22 is required before one 
gets close to the K/M dwarf members of this cluster assuming it does lie at 10-11~kpc as seems likely (see below). 
Note the faint limit of 2MASS is typically $K = 14.3$ and this only strictly applies to Galactic latitudes of $\vert$b$\vert$ $>$ 30~degrees (Skrutskie et al. 2006). At the low latitude of the cluster it could be up to 1.5~magnitudes brighter still.  We also note that the reddening vector and early type ($<$K5) stellar locus have similar (but not identical) slopes in the J-H versus H-K plot. Hence, the only objects likely to be seen with 2MASS data are possibly a few clump giants which are at the very limit of 2MASS. These stars almost lost in the background due to the large photometric errors and the crowding at 2MASS resolution in this compact cluster. 

\subsubsection{PN distance and reddening estimates}
Obtaining a reliable distance estimate for the PN itself  would have been more problematic until now.
Indeed, the determination of accurate distances to PNe is one of the more difficult observational problems in astronomy using current techniques. A few nearby objects have direct trigonometric parallaxes for their central stars (CSPN; Harris et al. 2006; Benedict et al. 2009), but the vast majority are far too remote for this technique to be applied.  A few resolved binary companions to CSPN have been used to provide spectroscopic parallaxes (Bond \& Ciardullo 1999; Ciardullo et al. 1999), but due to the wide range of temperatures and luminosities manifested in the CSPN  they are not suitable as standard candles, nor can their expanding PNe be used as standard rulers. Hence, any opportunity to derive nebular parameters for a PN at known distance, especially in a self-contained stellar system, is highly valuable as it enables precise determinations of the true physical size, luminosity, and ionised mass of the PN, as well as reliable parameters for the central star. 

We could already simply assume cluster membership for PHR1315-6555 given the other evidence.  However, we have also recently developed an innovative but simple 
H$\alpha$ surface brightness -- radius (SB-r) relation (Frew, Parker \& Russeil, 2006; Frew \& Parker 2006) that can independently provide a distance to this PN.  This simple but robust statistical relation has been built from over 200 critically evaluated PNe calibrators with reliable distances from primary techniques, and is accurate 
to $\sim$20--30 per cent if sub-trends are accounted for. It requires only an accurate integrated H$\alpha$ flux for each PN, a good reddening estimate and a precise measurement of the nebular angular size.

The reddening estimates towards the PN were derived  from careful evaluation and averaging of all our flux calibrated spectra and using the observed Balmer decrements. Note that care also needs to be taken with sky background subtraction which, with wide slits in particular, can affect the accuracy of the extinction values, especially during bright-time due to strong hydrogen absorption bands. Our spectra were obtained in dark sky or with a quarter moon below an airmass of 2. Fig. 2 does appear to reveal a low-level broad dip underlying the H$\beta$ emission lines but this broad feature does not seem to be present for either the H$\alpha$ or H$\gamma$ lines. We do not believe our extinction estimates are adversely affected. Interestingly,  the preliminary PN reddening estimate from the 2001 flux-calibrated SAAO spectral data, which covers the blue to red spectral region in a single exposure, was $E(B-V) = 0.71$ in very good agreement with the various cluster reddening values in the literature. Our later, higher quality PN spectra were obtained with the DBS on the ANU 2.3m which requires separate, careful flux calibration of the blue and red arms if Balmer decrement values are to be trusted (refer earlier comments). Our wide slit observations in May 2008 were also set at 19~arcseconds to encompass the entire PN and should, in principle, yield our best Balmer decrement values. The PN spectrum is not heavily contaminated by stars on the slit and the measured integrated [O~III] and H$\alpha$ fluxes are in excellent agreement with the MOSAIC-II results (see Table~\ref{PN-fluxes}).

Using our well determined and consistent H$\alpha$ flux, adopted average reddening value of $E(B-V) = 0.83$, and an accurate angular diameter from our high quality CTIO narrow band H$\alpha$ imaging we can then use the H$\alpha$ SB-r relationship presented in Frew, Parker \& Russeil (2006), which uses  the sub-trend applicable to bipolar PNe. 
Note the PN's angular size was measured at the 10 per cent isophote level (e.g. Kovacevic et al. 2010) to be consistent with the treatment of the calibrating sample for the H$\alpha$ SB-r relationship (Frew 2008).  Our completely independent distance estimate from this technique  is $10.5\pm 3.4$~kpc,  consistent with the adopted cluster distance of 10.4~kpc  (refer Table~\ref{summary}). This strongly supports PN cluster membership.  
Note that the sight-line to the PN and cluster at the Galactic longitude of 305~degrees is in the direction of the Sagittarius-Carina spiral arm so that spatial co-incidences between objects of widely different distances are possible so the distance agreement between PN and cluster is important.

\subsection{Galactic $\vert$$z$$\vert$ distance estimates for PN and cluster: a statistical argument}
Both PN and cluster are remote, with the open cluster being one of the most distant known from the Sun, so reddening approaching the asymptotic value along this sight line might be expected. If we assume a distance for both of $\sim10.4$\, kpc  as indicated earlier (see Table~\ref{summary}), 
then the PN and cluster sit around $\vert$$z$$\vert$$\sim$575~pc above the Galaxy's mid-plane. This is already  2-3 times the scale-height of the general Galactic PNe population (e.g. Zijlstra \& Pottasch 1991)  and approximately four times the scale-height found for intermediate age open clusters (Bonatto et al. 2006). We have checked the literature on the presence of the Galactic warp and $\vert$$z$$\vert$ offset at the Galactic longitude of the cluster and find that it is almost at a minimum at the cluster direction of $ l=305$~degrees (e.g. Momany et al. 2006 figs. 9-13; Levine et al. 2006) so the scale height arguments are not affected by any warp influence in this direction. In fact there is very little change in the Galactic warp signature around this longitude or with distance between 8 and 16\,kpc towards the cluster.
Within a 5-degree radius centred on the PN/cluster, i.e. within a region of 78~square degrees, there are only 36 known PN from the Acker (Acker et al. 1992) and MASH (Parker et al. 2006, Miszalski et al. 2008) catalogues and 43 known open-clusters from the Dias et al. (2002) open cluster catalogue. This yields surface densities of 
0.46 PN and 0.55 open clusters per square degree respectively in this region. Hence, having two unrelated sources within 20~arcseconds of each other on the sky that have $\vert$$z$$\vert$ distances much higher than their typical scale-heights would be statistically extremely unlikely, especially given the observed surface densities, but much more palatable if they were physically associated.

The  $\vert$$z$$\vert$ distance of ESO~96-SC04 is in itself interesting, and quite large for a Hyades-age cluster.  Indeed, the cluster was suggested by Frinchaboy et al. (2004) as a possible member of the Galactic Anticenter Stellar Structure (GASS), otherwise known as the Monoceros Stream (Pe\~narrubia et al. 2005).   This stream was first identified by Newberg et al. (2002) from Sloan Digital Sky Survey (SDSS) data.  Many literature studies have been done since, but no consensus yet exists regarding its nature and origin (e.g. see Hammersley \& Lopez-Corredoira 2010; Chou et al. 2010).  Alternatively, the SDSS data may represent the signature of a flared or warped disk in the outer Galaxy.   Nevertheless, the cluster age and metallicity is consistent with the proposed age-metallicity relation for the Monoceros Stream (see Figure~3 of Frinchaboy et al. 2004), though the age is considerably younger than any other system suggested as belonging to the stream.  In addition, the cluster velocity is consistent with GASS membership (see Frinchaboy et al. 2006a), but is also consistent with circular Galactic rotation at the adopted distance.  Eventually a proper motion determination of the cluster will confirm whether it is a potential physical member of the putative GASS.

\section{ Basic PN properties}

In this section we provide details of derived PN properties including abundances and other parameters derived from our spectroscopy.
If we assume an adopted cluster distance  (and hence PN distance) of 10.4\,kpc (Carraro et al. 1995) and use our measured  mean angular radius of 8~arcseconds  then this leads to an intrinsic radius of 0.4\,pc which is fairly typical of an evolved PN (Frew \& Parker, 2010).  We do not yet have detailed kinematic information for the main PN shell, so assuming a mean expansion velocity of 24 kms$^{-1}$, typical of an old PN (Frew 2008), this size corresponds to a nebular age of $\sim$11,000 years which again is well within anticipated PN lifetimes. A summary of the basic details for the PN, including position, size, morphology and age, is given as Table~\ref{basic}. 
Note the [O~III] absolute magnitude of the PN is M(5007)$\sim-$0.84 which is $\sim$3.6 magnitudes below the well-determined bright 
end cut off of the PN luminosity function, hinting at its evolved nature. 
The PN excitation class as determined from certain emission lines is E=7.8 following the scheme of Dopita \& Metheringham (1990), and Ex$\rho$=9.8 following the newly developed scheme of Reid \& Parker (2010) which appears to have a much tighter correlation with Zanstra temperature of the CSPN. Both values confirm the high excitation of the nebula as expected given the detection of HeII.
\begin{table}
\begin{center}
\caption{\label{PN-Cluster} {PN and cluster data comparisons }}
\begin{tabular}{lll}\\
\hline
Property &  PHR1315-6555 & ESO\,96-SC04 \\ \hline
Position RA (J2000) & $13^{h}15^{m}18.9{^s}$ & $13^{h}15{^m}16.0{^s}$\\
Position Dec (J2000) &  $- 65^{o}55'01''$ &  $-65{^o} 55'16''$\\
Position l,b & 305.368,   -3.158 & 305.362,   -3.162 \\
Distance (kpc) & $10.5\pm3.4$ & $10.4\pm1.8$ \\
Reddening   $E(B-V)$ & $0.83\pm0.08$  & $0.72\pm0.02$ \\
Radial velocity & $58\pm2.5$~kms$^{-1}$ & $57\pm5$~kms$^{-1}$ \\
\hline
\end{tabular}
\end{center}
\end{table}

\begin{table}
\begin{center}
\caption{\label{basic} Measured and derived properties for PN PHR1315-6555 (PN G305-3.-31). Note the cluster distance is used in the 
calculation of the parameters for the PN. }
\begin{tabular}{ll}\\
\hline
Characteristic   & Estimated value \\
\hline
Major \& Minor axes (max. extent) & $18\times14$~arcseconds\\
Major \& Minor axes (10\% contour) & $11.4\times10.2$~arcseconds\\
Morphology & Bipolar\\
Chemistry & Possible Type-I\\
DM excitation class & 7.8\\
RP excitation class & 9.8\\
Physical radius & 0.3~pc\\
Estimated age & 11,000~years\\
Galactic $z$ height below the plane & $\sim$575~pc\\
Estimated ionised mass & 0.5~M$_{\odot}$\\
$M_{5007}$ & $-0.8$\\
Estimated CSPN V mag & $23.5\pm1.0$\\
CSPN temperature (cross-over) & 209,000~K \\
CSPN temperature (RP Ex$\rho$ exc.class) & 265,000~K\\

\hline
\end{tabular}
\end{center}
\end{table}

\subsection{PN integrated line fluxes}
Integrated H$\alpha$, H$\beta$ and [O~III] line fluxes were first estimated from the measured fluxes from the narrow-slit SAAO observations in 2001 (knowing the dimensions of the slit) and by scaling up by a geometric factor to the full measured dimensions of the PN. These provided preliminary 
estimates.

\begin{figure*}
\includegraphics[width=18cm,height=2cm]{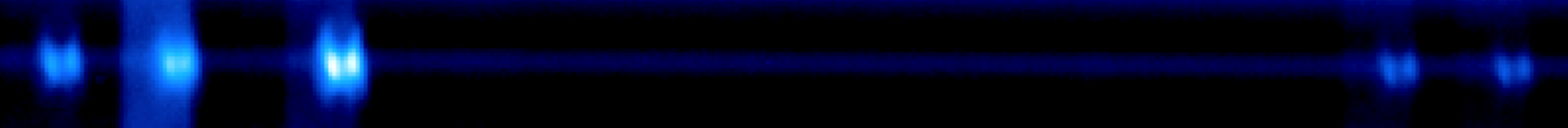}
\caption {2-D wide-slit spectral image of the 2.3m DBS red spectrum of PHR1315-6555 with wavelength increasing to the right. The image has been cleaned of cosmic rays and bias-subtracted. The bipolar nature of the PN is reaffirmed by this spectrum. The vertical direction is about 30~arcseconds and the slit width was 19~arcseconds. From left to right the lines are [N~II]6548\AA, H$\alpha$, [N~II]6584\AA~and the [S~II] doublet at 6717 \& 6731\AA~at the extreme right. Note how the PN emission lines are redshifted relative to the foreground wash of Galactic (not telluric) emission in this instance.} \label{DBS-red-wide}
\end{figure*}

Very wide-slit (19~arcsecond) 2.3m observations were subsequently taken with the ANU 2.3m telescope in May 2008 with the DBS  to obtain more reliable integrated line fluxes for the entire PN for the most important lines. A 2-D image of the  DBS red image with the 1200R grating is shown in Fig.~\ref{DBS-red-wide} which clearly shows we have observed the entire nebula and also re-affirming the PN's bipolar morphology. Appropriate flux standards were observed with a 15~arcsecond wide slit and the data reduced using standard iraf tasks 
supplement by the PNDR package developed by one of us (BM)\footnote{http://star.herts.ac.uk/$\sim$brent/pndr/}. 

Average measured fluxes F($\lambda$) for  detected PN lines from our deep  narrow and wide-slit flux-calibrated spectra are given  in Table~\ref{pnfluxes} normalized to H$\beta$ = 100.  These lines were dereddened according to the Howarth (1983) extinction law and are given in the third column as I($\lambda$). The observed logarithmic extinction $c$  was  an average value determined from our observed Balmer decrements  in our  flux-calibrated spectra.  We obtain $c_{\beta}$ = 1.21 or E(B-V) = 0.83, which is in good agreement with the average cluster reddening determination  given the errors.

Finally, accurate, integrated flux estimates were also made from the MOSAIC-II camera narrow-band imaging 
in both [O~III]  and H$\alpha$ obtained in June 2008. The H$\alpha$ flux was determined  after de-convolving the  [N~II] 
contribution established from our spectroscopy and after modifying line values by the respective filter transmissions  at their velocity-shifted wavelengths. These direct photometric  values  agree completely,  to within the errors, to  estimates  
derived from our  flux-calibrated wide-slit spectroscopy. These three independent sets of integrated line flux estimates
are presented in Table~\ref{PN-fluxes} and are shown to be in excellent agreement to within the errors. 
\begin{table}
\begin{center}
\caption{\label{PN-fluxes} {Comparison of key independent PN line flux estimates}}
\begin{tabular}{lccc}\\
\hline
Observation &   Date & log\,F~H$\alpha$ & Log\,F~[OIII] \\ \hline
SAAO narrow-slit & May 2001 & $-12.49\pm0.10$ & $-12.28\pm0.10$ \\
DBS wide-slit & May 2008 &$ -12.44\pm0.05$ & $-12.32\pm0.05$ \\
CTIO MOSAIC-II  & June 2008 & $-12.45\pm0.02$ & $-12.35\pm0.02$\\
\hline
\end{tabular}
\end{center}
\end{table}

\begin{table}
\begin{center}
\caption{Average line fluxes and flux ratios for PHR1315-6555 as measured from our flux calibrated spectra as summarised in Table~\ref{specsum}. The  I($\lambda$) values have been corrected using the reddening law of Howarth (1983) and were input to the {\sc hoppla} plasma  code for subsequent abundance estimates. {\bf Note the line ratios represent average values from our various flux calibrated spectra. The plasma diagnostics at the end, including extinction, are formal {\sc hoppla} outputs based on the input line values.}}
\label{pnfluxes}
\begin{tabular}{llrrrr}
\hline
ID & $\lambda$ & F($\lambda$)  & I($\lambda$)\\
 \hline
{[\rm O \sc ii]} 		& 3727 & 163 	& 333\\
{[\rm Ne \sc iii]} 		& 3869 & 71 	& 135 \\
H$\gamma$ 		& 4340 & 28 	& 40\\
{[\rm O \sc iii]} 		& 4363 &  20 	& 28  \\
He {\sc ii} 			& 4686 & 62 	& 70\\
H$\beta$ 			& 4861 & 100 	&  100\\
{[\rm O \sc iii]} 		& 4959 &  	307	& 287 \\
{[\rm O \sc iii]} 		& 5007 & 944 	& 854\\
{[\rm N \sc i]} 		& 5200 & 16 	& 13\\
{[\rm N \sc ii]} 		& 5755 & 14 	& 8.5\\
He {\sc i} 			& 5876 & 20 	& 11 \\
{[\rm O \sc i]} 		& 6300 & 49 	& 22\\
{[\rm S \sc iii]}		& 6312 & 13 	& 5.9\\
{[\rm O \sc i]} 		& 6364 & 16 	& 7.1 \\
{[\rm N \sc ii]} 		& 6548 & 359 	& 148 \\
H$\alpha$ 		& 6563 & 667 	& 272\\
{[\rm N \sc ii]} 		& 6584 & 1115 & 453\\
He {\sc i} 			& 6678 & 6 	& 2.4\\
{[\rm S \sc ii]} 		& 6717 & 127 	& 49\\
{[\rm S \sc ii]} 		& 6731 & 104 	&  40\\
He {\sc i} 			& 7065 & 6 	& 2\\
{[\rm Ar \sc iii]} 		& 7135 & 58 	& 19\\
{[\rm O \sc ii]} 		& 7320 & 15 	& 4.7\\
{[\rm O \sc ii]} 		& 7330 & 10 	& 3.1\\
\hline
$[$N~II$]$/H$\alpha$ &  2.21 $\pm$ 0.19 	&  & \\
5007/H$\beta$ 		&  9.44 $\pm$ 0.10 	&   &\\
$[$S~II$]$ 6717/6731 &  1.22  $\pm$ 0.09 	&  & \\
$[$S~II$]$/H$\alpha$ &  0.35 $\pm$ 0.02 	&   &\\
\hline
$c$H$\beta$ 		&  1.21 	& &  &  &\\
N$_{e}$ 			& 160 		&&& &\\
T$_{e}$ (average) 	& 18100\,K &&&&\\
\hline \\
\end{tabular}
\end{center}
\end{table}

\subsection{PN chemical abundances and ionised mass}

Our collection of optical spectra for the PN show  enrichments of He and N, i.e. this is evidence for a Type~I PN (Peimbert \&
Torres-Peimbert, 1983; Kingsburgh \& Barlow, 1994). Such PNe are thought to derive from higher mass progenitor stars. This is consistent with the estimated cluster main sequence turn off mass of  $\sim2.5M_{\odot}$. Abundances from such PNe for unprocessed elements (e.g. O and S) as well as processed elements (C
and N) directly yield the enhancements that occurred during AGB evolution. If the progenitor star is indeed at 2.5~M$_{\odot}$ 
then this would imply that the lower mass limit for hot bottom burning is a little lower than predicted by existing stellar evolutionary
models. This would be a very important new constraint for theory.

A preliminary abundance analysis of the flux-calibrated SAAO spectrum indicated likely Type~I chemistry but the S/N was too low for some of the crucial lines to be more quantitative. Much deeper nebular spectra were obtained with the 2.3m DBS in May 2006 and May 2008.  Despite the fact that a dichroic splits the light into the separate blue and red arms of the DBS, observations of spectrophotometric standard stars and the carefully flux-calibrated spectra naturally enable a combined red and blue spectrum to be used, as shown in Fig~\ref{PN-spec}. We have confidence in this process because the integrated fluxes for the [O~III] and H$\alpha$ lines are in excellent agreement with our CTIO narrow-band imaging (see Table~\ref{PN-fluxes}). We note that the estimated PN oxygen abundance is 0.4~dex lower than solar but the sulphur appears to agree with neon but not with oxygen so the metallicity is not yet well constrained.
Also note that weak H$\gamma$ absorption from a star on the slit has slightly depressed the observed emission. This helps to make the observed H$\gamma$ to [O~III] 4363\AA~ratio in Fig~\ref{PN-spec} even closer to that typically seen in symbiotic stars (which is clearly not the case here). This should not affect our abundance estimates.

The observed dereddened I($\lambda$) line fluxes were analysed with the {\sc hoppla} plasma diagnostics code (Acker et al. 1989) which  performed the abundance analysis  
and applying the Howarth (1983) reddening law and the usual ionization correction factors for unseen stages of ionization. The results are given in Table~\ref{abund}. From these results the PN may be classified as a Type~I object  according to the definition of Peimbert \& Torres-Peimbert (1983), but is on the cut-off for Type~I used by Kingsburgh \& Barlow (1994) so the Type~I nature is not absolutely established though certainly possible. 
\begin{table}
\begin{center}
\caption{Elemental abundances by number for PHR1315-6555 from {\sc hoppla}, given in the usual notation of 12 + $\log (n({\rm X})/n({\rm H}))$.  For comparison, the abundances for Type~I and non-Type~I PN are taken from Kingsburgh \& Barlow (1994) and solar abundances from Asplund et al. (2005).}
\label{abund} 
\begin{tabular}{lcccc}\\
\hline
Element   & PHR1315--6555 & Type~I & non-Type I & Solar\\
\hline
      He        &	11.18 	& 	11.11 	&11.05 	&10.93\\
      N           &  	8.20 		&      	8.72		& 8.14	& 7.78 \\
      O           &  	8.26 		&    	 8.65 	& 8.69 	& 8.66 \\
      Ne	 &  	7.73   	&        8.09     	& 8.10  	 & 7.84\\
      S           &  	6.97  	&   	6.91 		& 6.91 	& 7.14 \\
      Ar          &  	5.80   	&    	6.42 		& 6.38 	& 6.18    \\
\hline
log${\rm (N/O)}$    	&      $-0.06$   	 & $+0.07$ 	& $-0.55$ &	$-0.88$  \\
log${\rm (Ne/O)}$    	&      $-0.53$   	 &   $-0.56$       	& $-0.59$  &	$-0.82$  \\
log${\rm (S/O)}$    	&      $-1.29$   	 & $-1.74$		&  $-1.78$ &	$-1.52$  \\
log${\rm (Ar/O)}$    	&      $-2.46$   	 & $-2.23$ 	&  $-2.31$ &	$-2.48$  \\
\hline
\end{tabular}
\end{center}
\end{table}

Following Pottasch (1996), the ionised mass of a PN can be estimated  by:
\begin{equation}
\label{eq:pott}
M_{\rm ion} = 4.03 \times 10^{-4} \epsilon^{1 \over 2} d^{5 \over 2} F(H\beta)^{1 \over 2} \Theta^{3 \over 2} M_{\odot}
\end{equation}
where $d$ is the distance to the PN in kpc, F($H\beta$) is the reddening-corrected flux in units of 10$^{-11}$ \ergcms, and $\Theta$ is the mean PN 
radius in arcseconds.  We assume a filling factor $\epsilon$ = 0.3 (e.g. Pottasch 1996; Pierce et al. 2004) and estimate an ionised mass of $\sim
$0.5\,M$_{\odot}$. 

From our spectral observations the PN is likely to be optically thick (e.g. Kaler \& Jacoby 1989) as our spectra exhibit strong
emission from  [O~II], [N~II], [S~II], simultaneously with strong He~II emission, and the weak detection of [O~I]~6300 \& 6364\AA~and [N~I]~5200\AA. 


\section{Estimated PN central star properties}
Any derivation of the central star parameters such as progenitor mass, temperature and magnitude, makes an explicit assumption that the PN does not have a binary central star. Recently Miszalski et al. (2009a) have shown that only 17$\pm$5 per cent of all PNe have a close binary CSPN while Miszalski et al. (2008b; 2009b) have also recently found additional evidence for close binary stars in canonical bipolar nebulae (e.g. M\,2-19) as previously noted by Bond \& Livio (1990) and Frew \& Parker (2007).  Consequently we  intend to obtain deep I-band exposures  to exclude the possibility of a cool companion though a double-degenerate system would still remain a possibility. Nevertheless, in the absence of any current evidence for binarity, a single CSPN is a reasonable default assumption.

\subsection{CSPN Magnitude and Temperature}

On our off-band nebula images, there are more than a dozen stars superposed on the body of the PN.  There is a possible candidate central star within 3~arcseconds of the poorly-defined centre of the PN on the CTIO [O~III]-off band image.  To confirm this, we retrieved a V-band image from the ESO archive (proposal CI G. Carraro, May 1994) which clearly showed the star; this image was used in the study of Carraro et al. (1995).   This star was also detected by Carraro et al. (2005) as star no. 1385, and they measured V = 19.77, $B-V$ =  0.49, $V-I$ =  0.23.   Using E(B-V) = 0.83, the dereddened colours are $(B-V)_{0}$ = $-0.34$ and $(V-I)_{0}$ =  $-0.75$ which is consistent with a hot star. 

However, using our own accurate spectrophotometric data of the PN shell, we predict an apparent V magnitude of $23.5\pm1$ using the crossover (Ambartsumyan) method (Kaler \& Jacoby 1989) and our available dereddened line fluxes.  This method calculates a temperature (and CSPN magnitude) by forcing agreement between the H and He Zanstra temperatures from the nebula, and necessarily assumes that the nebula is optically thick (as already indicated in section 5.2).  We derive $T_{\rm cross} = 209,000$\,K which makes the CSPN one of the hottest known.   A lower quality CSPN  $T_{eff}$  estimate can also be made based on the new PN excitation class parameter established by Reid \& Parker (2010). This new excitation class formulation appears to have a far more direct correlation with the best determined Zanstra temperatures for central stars (see their Fig. 16).  If their new Ex$\rho$ evaluation for the PN of 9.8 is used in their empirical transformation of this value to a 
Log~$T_{eff}$ then we obtain $T_{Ex\rho} = 265,000$\,K, an even more extreme value.
Unfortunately, neither of these values are  accurate enough to determine a reliable mass for the CSPN from the theoretical HR diagram though a high CSPN temperature does seem to be indicated.  Such high temperatures appear to be incompatible with the assumed nebula age as they are reached only by central stars more massive than 0.7M$_{\odot}$ (using Bl\"ocker evolutionary tracks) but none of the models have such high temperatures after 11,000 years (the estimated PN age). In these models for such massive central stars the stars
evolve very quickly and by the estimated age of the host PN will now be well down on the cooling track. If we assume that the luminosity and PN age estimates are reasonable and if we assume a  typical residual core mass of  0.6-0.65~M$_{\odot}$ then $T_{eff}$ should be in the range
 100,000-140,000\,K less than that implied from our other estimates. We hope that better data will help resolve this inconsistency.
 
If the star visible on our images is the CSPN, we find a reddening-corrected absolute magnitude of $M_{V}$ = +2.1 and using the integrated 
H$\alpha$ flux we measured for this optically thick nebula, we determine a He~II Zanstra temperature of 95,000~K.  However, this is much cooler than obtained using the cross-over method, and it may be that we have identified a lightly reddened foreground star, and not the true CS which is predicted to be considerably fainter.  Alternatively, we may have identified an unresolved binary central star in this PN.  It is important to positively identify the true CSPN, but for this we will require deep 8-m or HST photometry of sufficient S/N to identify it unambiguously against the background nebula in such a crowded field.

\subsection{The progenitor mass of PHR1315-6555 and the IFMR} 
Accurate photometry of the central star is essential to better determine the central star luminosity and temperature via the Zanstra method.
Fitting these parameters to the theoretical HR diagram will provide a central star mass which will be an important additional datum for the
white dwarf IFMR (e.g. Dobbie et al. 2009), assuming it is a single star. This region of the IFMR  (e.g. progenitor masses of M$\sim$2.5M$_{\odot}$) is currently sparsely populated, though we note the datum for PN NGC~7027 which does fall in this region (Zijlstra et al. 2008).

A key-point of interest here is the apparent sub-solar oxygen abundance assuming the PN is a cluster member. 
An independent but very preliminary cluster metallicity estimate is also given by Frinchaboy et al. (2006b) of [Fe/H]=-0.51. This was based on only
two stars however, and although consistent with their membership of an old cluster, should be treated with caution. 
The cluster also has a similar age to the Hyades and Praesepe clusters (600--800 Myr), which hold a total of 15 apparently single,
non-magnetic white dwarfs which have been used to constrain the IFMR at [Fe/H]=+0.14.  If the Frinchaboy (2006b) cluster metallicity estimate is correct, then ESO 96-SC04 may be metal poor by a factor 4--5 with respect to these populations and thus offers the chance to probe a metallicity regime which has not yet been well explored at this age. The fact that the IFMR is 
metallicity dependent is well known from earlier AGB/PN work (e.g. Vassiliadis \& Wood 1993) and confirmed more recently by the work of Marigo \& Girardi
(2007) though the effect may be less pronounced  around progenitor masses of M$\sim$2.5M$_{\odot}$.
Nevertheless, this object could provide an additional probe of such an effect.

\section{Conclusions}
We have found a faint, bipolar,  high-excitation planetary nebula (PHR1315-6555) of possible Type~I chemistry in the intermediate-age open cluster ESO\,96-SC04,  based on our initial discovery from the AAO/UKST H$\alpha$ Survey and subsequent confirmatory  spectroscopy.  The PN was missed on earlier broadband CCD imaging studies of the cluster as it is compact and of relatively low surface brightness.  We have several key arguments and other corroborating strands of evidence that together present an extremely compelling case for a physical association between the PN and the cluster. These arguments comprise very close angular separation of PN to the compact cluster core, excellent agreement of PN and cluster radial velocities, compatible, independent distance determinations to PN and cluster, consistent reddening estimates,  
and  Galactic scale height arguments. Finally, the estimated PN physical properties are consistent with the cluster turn-off mass and distance. PHR1315-6555 is currently the only bona-fide PN known to be unequivocally associated with an open cluster in the Galaxy.  The importance of this physical association to the WD initial-to-final mass relation is stressed.  Follow-up observations are planned to unambiguously identify and measure the CSPN and address the question of its possible binarity.

\section*{Acknowledgements}
This study used data from the AAO/UKST H$\alpha$ Survey, produced by the Anglo-Australian Observatory and  Particle Physics and Astronomy Research Council (UK).    This research made us of the SIMBAD database and Vizier service, operated at CDS, Strasbourg.  QAP, AK and DJF acknowledge the support of ANSTO to enable them to undertake the spectroscopic and imaging observations. AK wishes to acknowledge Macquarie University for a PhD scholarship.
We are grateful to the South African Astronomical Observatory, MSSSO, ATAC and NOAO for the generous award of observing time.  We thank the referees David Turner and Albert Zijlstra for useful suggestions on earlier drafts of this manuscript.



\end{document}